\documentstyle[12pt]{article}
\newcommand{\be}{\begin{equation}}
\newcommand{\ee}{\end{equation}}
\newcommand{\ba}{\begin{eqnarray}}
\newcommand{\ea}{\end{eqnarray}}

\begin{document}
\renewcommand{\baselinestretch}{1.3}
\small\normalsize
\renewcommand{\theequation}{\arabic{section}.\arabic{equation}}
\renewcommand{\thesection}{\Roman{section}.}
\language0

\title{Spectral properties and lattice-size dependences in cluster algorithms}
\author{Werner Kerler}
\date{\sl Fachbereich Physik, Universit\"at Marburg, D-35032 Marburg,
Germany}
\maketitle
\begin{abstract}
Simulation results of Ising systems for several update rules, observables, and
dimensions are analyzed. The lattice-size dependence is discussed for the
autocorrelation times and for the weights of eigenvalues, giving fit results in
the case of power laws. Implications of spectral properties are pointed out and
the behavior of a particular observable not governed by detailed balance is
explained.
\end{abstract}

\newpage

\section{Introduction}
\hspace{0.35cm}
The reduction of critical slowing down in Monte Carlo simulations achieved
by cluster algorithms \cite{sw} has provided access to particularly interesting
regions of physics. So far, however, efficient algorithms of this type are only
known for limited classes of systems. For example, extensions are particularly
desirable to gauge fields. There are treatments \cite{bh} of the discrete case
of $Z_2$ gauge theory in 3 dimensions where one has duality to the Ising
model. However, for continuous gauge theories only for the very special
situation of one time step in $SU(2)$-theory at finite temperature a
successful algorithm using an Ising embedding has been obtained \cite{bems}.

A prerequisite for any further developments of cluster algorithms is a better
understanding of the details which are responsible for specific properties. To
make progress within this respect recently \cite{Ke} the effects of flipping
rules have been studied. In the present paper more details of the simulation
results are presented and analyzed. In addition theoretical discussions and
explanations of some observations are given.

Sec.~II contains an overview of the simulations of Ising systems and of the
autocorrelation fits. It also presents ratios of main modes, quantitative
results about fast modes and details of the dependence of the cluster-size
observable on lattice size. In Sec.~III. the lattice-size dependences of the
autocorrelation times and of the weights of eigenvalues are given and
discussed.
The analysis is based on power law fits wherever this is possible. In other
cases, where also logarithmic fits do not work, the considerations rely on
figures.

Effects of spectral properties are discussed in Sec.~IV. It is pointed out that
the results for the spin-dependent observables conform with the theoretical
expectations. The peculiar behavior of a particular observable, describing the
sum of sizes of the clusters with flipped spin, is explained in Sec.~V. It
provides an example of a situation where there is no detailed balance and also
of the ways to deal with more general transition probabilities.

\section{Monte Carlo simulations and results}\setcounter{equation}{0}
\hspace{0.35cm}
The Ising systems have been studied at the infinite-volume critical point on
hypercubic lattices with periodic boundary conditions in 2, 3, and 4
dimensions. The $\beta$-values used are log$(1+\sqrt{2})/2$~, 0.221650
\cite{bptr}, and 0.149663 \cite{gsm}, respectively.
The observables measured in all cases are the energy $E$, the magnetization
$M$ , the susceptibility $\chi$ , and $C$, the relative size of the clusters
the spins of which are flipped ( i.e.~the number of flipped spins divided by
the number of sites). For each of the observables also the autocorrelations
have been determined. Relatively high statistics has been collected (with
$2\times10^6$ to $10^7$ sweeps for lattice sizes up to $64^2$, $16^3$, and
$9^4$, and up to $5\times10^5$ sweeps for some larger lattices) in order to
make the fits to the autocorrelation functions possible.

In addition to flip probabilities according to Swendsen and Wang \cite{sw} (SW)
and to Wolff \cite{wo} (WO), and of flipping the largest cluster \cite{bc}
(LC),
a prescription \cite{Ke} (SC) has been investigated which requires to flip the
largest cluster and those not in contact with it. Contact between two clusters
means that there is at least one site belonging to the first cluster which is
only one lattice spacing away from a site belonging to the second one.

The occurring time scales and the fit procedure for the autocorrelation
functions have already been discussed in detail in \cite{Ke}. With respect to
the main modes of the transition spectrum it turned out that SC and SW are
described by functions of form $ce^{-t/\tau}$ and LC and WO by functions of
form $ce^{-t/\tau}+c_2e^{-t/\tau_2}$ (representing terms in the normalized
autocorrelation functions).  In addition to the values of $\tau$, $c$,
$\tau_2$, and $c_2$ also the integrated autocorrelation times $\tau_{int}^E$,
$\tau_{int}^{\chi}$, $\tau_{int}^M$, and $\tau_{int}^C$ have been determined.

Presenting more details it is to be noted that the ratios $\tau/\tau_2$ in
$d=2$, 3, and 4, respectively, have the values 3.2(4), 1.9(2), and 1.8(2) for
LC and 2.7(2), 2.3(2), and 2.5(2) for WO. These numbers in $d=3$ and 4 apply
to the whole range considered while in $d=2$ the numbers at $L=32$ are given.
This is related to the fact that in $d=2$ the lattice-size dependence is not
a power law as will be discussed in Sec.~III.

In addition to the main modes included in the fits, fast modes only noticeable
in the vicinity of $t=0$ occur. In the following to give quantitative results
on their magnitudes the overall coefficients $c_f=1-\eta(0)$ are used where
$\eta(t)$ is the fit function determined at larger $t$ and having one of the
forms mentioned above.

In general the $c_f$ slightly increase with lattice size and
dimension. In the case of SW the $c_f$ are in the range of about 0.03 to 0.07
for $E$, of about 0.05 to 0.2 for $\chi$, and of about 0.1 to 0.3 for $M$. For
the decay time of a fast mode a typical number (obtained from a 4-parameter fit
for $d=4$ and $L=8$) is 0.7(2) as compared to 4.79(3) of the leading mode.
Estimates in other cases give similar results. For $C$ only a fast mode
concentrated at $t=0$ is observed within errors (with an upper limit of
$10^{-3}$ for other contributions), i.e.~ $\rho(t)=\delta_{t,0}$ is found. In
the case of WO the behavior of fast modes is qualitatively the same as for SW
for $E$, $\chi$, and $M$. For $C$ again a strong mode proportional to
$\delta_{t,0}$ occurs, however, also the main modes still contribute.

In the case of SC fast modes show rather different behaviors for different
observables. For $E$ there are no or almost no fast modes within errors. For
$M$ the overall coefficients $c_f$ are in the range of about 0.2 to 0.3 and one
observes oscillations, i.e. contributions with a time dependence of the form
$(-1)^te^{-t/\tau_{\nu}}$. For $\chi$ the $c_f$ are in the range of about 0.1
to
0.2 and decaying as well as oscillating contributions occur. For $C$ in $d=2$
the coefficients $c_f$ are rather accurately known (being $1-c$ with the fit
result for $c$ in Table 2). With lattice size they increase strongly from
negative to positive values. The time dependences observed change from
oscillations at small lattices to a purely decaying behavior at larger
lattices.
For LC the observed fast modes are very similar to those seen for SC.

For LC and WO the measured values of $\langle C\rangle$ decrease with $L$
according to power laws. Results from fits to $kL^z$ for these cases are given
in Table 1.  For SW one has, of course, $\langle C\rangle =\frac{1}{2}$. For SC
after decreasing with $L$ a minimum is reached which occurs earlier in higher
dimension. This behavior reflects the fact that the sum of the relative sizes
of
the clusters not in contact with the largest one increases. To describe the
situation within this respect in detail in Fig.~1 $\langle C\rangle_{LC}$ and
the difference $\langle C\rangle_{SC}-\langle C\rangle_{LC}$ denoted by DC are
shown. The fast increase of DC at larger $L$ in higher dimension makes clear
that one should account for connectivity (as discussed in \cite{Ke}) also among
clusters not in contact with the largest one.

\section{Lattice size dependences}\setcounter{equation}{0}
\hspace{0.35cm}
For the quantities $\tau$, $c$, $\tau_2$, $c_2$, $\tau_{int}^E$,
$\tau_{int}^{\chi}$, $\tau_{int}^M$, and $\tau_{int}^C$ obtained in the fits to
the autocorrelation functions next the dependence on the lattice size $L$ is to
be considered. Fits to the law $kL^z$ have been performed for all of those
quantities whenever this has been possible within errors.  Results of these
fits
are given in Tables 2 and 3 (omitting some less interesting or less accurate
cases).

In these power-law fits again interval stability and magnitudes of
$\chi^2$-values have been checked. Throughout this has led to exclude $L=4$
(though graphically the deviations mostly appeared small). Systematic errors
have been estimated in particular from residual interval instabilities. They
are smaller than the statistical errors in the one-mode cases and of the
same order in the two-mode cases. While in the few examples of power-law fit
data already given in \cite{Ke} the errors are only statistical ones, here the
estimated systematic errors have been combined with the statistical ones to
obtain the errors presented in the tables. Further, some of the numbers have
been conservatively rounded to allow an easier overview.

The autocorrelation times in $d=3$ and $4$ within errors are consistent with
power laws. For $d=2$ in general different dependences on the lattice
size $L$ are found for all flipping rules. In such cases the symbol $\cap$ is
entered in Tables 2 and 3 if they are listed at all. For SC and SW in $d=2$
logarithmic fits for $\chi$ and $M$, represented as power law fits of
exp($\tau_{int}$) in Table 2, are possible while they can be ruled out for $E$
and are also not possible in other cases.

Due to the accuracy of the data it is to expected that the different behaviors
of the autocorrelation times in $d=2$ will persist at higher $L$. On the other
hand, usually one expects only powers and logarithms at sufficiently large $L$.
How this is to be reconciled with the numerical results is at present an open
question. Thus data on various autocorrelation times for SC, LC, and WO which
could not be described by fits are presented in Figs. 2, 3, and 4.

The results on the lattice-size dependences of the weights of the eigenvalues
are given by the numbers for the coefficients in Tables 2 and 3 and in
Figs.~5 -- 7. It is seen that depending on dimension and particular observable
decreasing, constancy, or increasing is observed. The changes of the
coefficients with $L$ in general are not large such that some effort has been
necessary to get them. Generally the change is slightly stronger for $M$ than
for $\chi$ while for $E$ it is small. For $C$ the behavior is frequently rather
different from that found for the other observables.

For SC the fits show (Table 2) that with increasing $d$ negative
slopes turn into positive ones. For SW generally decreasing behaviors occur,
with power laws (Table 2) and with deviations from this in $d=2$ (Fig.~5).
For LC and WO at least in $d=3$ and $4$ (and apart from the observable $C$)
the general tendency is that $c$ increases and that $c_2$ decreases as is
indicated to some extend in the examples of Figs. 6 and 7. The sum $c+c_2$,
on the other hand, remains essentially constant (Table 3).

For the coefficients deviations from a power law may be caused by the
self-adjointness of the transition matrix, which implies that $c$ cannot become
larger than 1 as will be pointed out in Sec.~IV. Deviations of this type are
shown in Fig. 5 for the example of SW in $d=2$. On the other hand, if there is
no such restriction $c$-values larger than 1 occur. That they do is seen in the
example of SC for $C$ in d=2, the power-law data of which are given in Table 2
and the mechanism for which is explained in Sec.~V.

It is to be noted that (at least at the values of $L$ which can be reached in
practice) simultaneous power-law behaviors of $\tau_{int}$, $\tau_{\nu}$, and
$c_{\nu}$ cannot be expected in general. The reason for this is that a term
$c_{\nu}e^{-t/\tau_ {\nu}}$ from $\rho(t)$ gives the contribution $c_{\nu}
(e^{1/\tau_{\nu}}-1)^{-1}$ to $\tau_{int}= \frac{1}{2}+\sum_{t\ge 1}\rho(t)$.
Only if the decay time of the leading mode is large and its coefficient is not
small, such that the approximation $\tau_ {int}\approx c\tau$ holds, the laws
$k_cL^{z_c}$ and $k_{\tau}L^{z_{\tau}}$ combine to the simultaneous law
$k_c k_{\tau} L^{z_c+z_{\tau}}$. The data in Tables 2 and 3 are in agreement
with these considerations. Empirically there is some indication that primarily
one has power behavior of $\tau$ and of $c$ while for $\tau_{int}$ one relies
on the decribed approximation.

\section{Effects of spectral properties}\setcounter{equation}{0}
\hspace{0.35cm}
The spectral properties of the transition matrix $W(\sigma;\sigma')$ manifest
themselves in the autocorrelation function
\be
R(t)=\langle f_s f_{s+t}\rangle-\langle f_s\rangle^2
\label{4e10}
\ee
where with the probability distribution $\mu(\sigma)$ and an observable
$f(\sigma)$ one has
\be
\langle f_t\rangle=\sum_{\{\sigma\}}\mu(\sigma)f(\sigma)
\label{4e54}
\ee
and
\be
\langle f_t f_{t+1}\rangle=\sum_{\{\sigma\},\{\sigma'\}}
\mu(\sigma)f(\sigma)W(\sigma,\sigma')f(\sigma') \quad .
\label{4e55}
\ee
To discuss these spectral properties a Hilbert space with an inner product
defined by
\be
(f,g)= \sum_{\{\sigma\}}\mu(\sigma) f^*(\sigma)g(\sigma)
\label{3e23}
\ee
is convenient \cite{msls}. In the indicated metric detailed balance is just
the condition for self-adjointness.

Using the spectral representation $\sum_{\nu}\lambda_{\nu} F_{\nu}$ of the
transition operator with projectors $F_{\nu}$~, eigenvalues $\lambda_{\nu}$~,
and $\lambda_1=1$~, one gets
\be
R(t)=\sum_{\nu\ne 1}\lambda_{\nu}^{\,t}\:(f,F_{\nu}f) \quad .
\label{4e11}
\ee
{}From stationarity and ergodicity it follows that apart from the eigenvalue 1
all
eigenvalues of the transition matrix are in the interior of the unit circle. If
detailed balance holds the self-adjointness of the transition operator implies
real eigenvalues and positive weights. Then the parametrizations
$\lambda_{\nu}= e^{-1/\tau}$ and $=-e^{-1/\tau}$ are appropriate which give
the time dependences $e^{-t/\tau}$ and $e^{-t/\tau}(-1)^t$, respectively.

The spectral representation of the normalized autocorrelation function
$\rho(t)=R(t)/R(0)$ has the form
\be
\rho(t)=\sum_{\nu}\lambda_{\nu}^{\,t}\:c_{\nu} \quad .
\label{4e31}
\ee
{}From this and $\rho(0)=1$ one gets the condition $\sum_{\nu}c_{\nu}=1$ for
the
coefficients. Thus, if detailed balance holds, in which case all $c_{\nu}$ must
be positive, they also cannot exceed 1~. Numerical results illustrating this
have been discussed in Sec.~III.

In cluster algorithms the conditioned probability $A(\sigma;n)$ for getting the
bond configuration $\{n\}$ if the spin configuration $\{\sigma\}$ is given is
important. The joint probability distribution $\mu_J(n,\sigma)$, the occurrence
of which in the case of the SW rule has been pointed out in \cite{es}, can be
generally expressed by $\mu_J(n,\sigma)= \mu(\sigma)A(\sigma,n)$. For the SW
flipping rule, where the new cluster spins are choosen independently of the old
ones, the conditioned probability to get the new configuration $\{\sigma'\}$ if
the configuration $\{n\}$ is given may be written as
\be
\tilde{A}(n;\sigma)=\mu_J(n,\sigma)\left(\sum_{\{\sigma\}}\mu_J(n,\sigma)
\right)^{-1}
\label{2e10}
\ee
and the transition matrix gets the special form
\be
\tilde{W}(\sigma;\sigma')=\sum_{\{n\}} A(\sigma;n)\tilde{A}(n;\sigma') \quad .
\label{2e15}
\ee
For (\ref{2e15}) detailed balance follows straightforwardly from the
definitions (provided that $A$ has the appropriate properties, which is
guaranteed by the usual construction). Further from $(f,\tilde{W}f)=
(V_J f,V_J f)$ with $V_J(n,\sigma,n',\sigma') = \tilde{A}(n; \sigma')
\delta(n,n')$ it is seen that the transition matrix is positive in this case.
This conforms with the observations of a real positive spectum in the
simulations for SW.

For the other flipping rules considered the more general transition probability
\be
W(\sigma;\sigma')=\sum_{\{n\}} A(\sigma;n) B(n,\sigma;\sigma')
\label{2e11}
\ee
applies, where $B$ is the conditioned probability to arrive at $\{\sigma'\}$ if
both $\{\sigma\}$ and $\{n\}$ are given. The choice of $B$ is important for the
spectral properties. Now (in addition to the appropriate properties of $A$) $B$
must be symmetric in $\sigma$ and $\sigma'$ in order that (\ref{2e11})
satisfies
detailed balance. For the other rules considered detailed balance follows
readily on the indicated basis. Thus the spectrum is again real, however, no
longer necessarily positive. In fact, for SC and LC oscillating contributions
related to negative eigenvalues are observed.

Because the factor $\sqrt{2\tau_{int}}$ enters the statistical error and the
integrated autocorelation time is related to $\rho$ by
\be
2\tau_{int}= 1+2\sum_{t\ge 1}\rho(t)
\label{4e31a}
\ee
in view of (\ref{4e31}) it should be obvious that not a positive spectum as in
the SW case but rather a negative one is optimal. If this could be realized a
drastic reduction of the errors by anticorrelations would be obtained.

The general law has been found \cite{Ke} that the
leading eigenvalue of the transition matrix tends to dominate if the pattern of
clusters subject to spin flips becomes extended. The indicated relation between
flipping pattern and spectral distribution is reminiscent of familiar relations
for conjugate variables in Fourier transforms. The reason for this is that the
spectral represention of the transition matrix relates changes of
configurations
to the spectrum in an analogous way. This explains the observation.

The discussion of spectral properties on the basis of the scalar product
(\ref{3e23}) for observables of type $f(\sigma)$ applies to $E$, $\chi$, and
$M$.  For the quantity $C$ more general considerations are necessary which will
be outlined in Sec.~V.

\section{Joint probabilities and spectra}\setcounter{equation}{0}
\hspace{0.35cm}
Because configurations $\{\sigma\}$ as well as configurations $\{n\}$ are
generated, in addition to observables depending on $\sigma$ one may also
measure more general ones depending on $\sigma$ and $n$. One then could
consider
transitions with respect to the joint distribution $\mu_J(\sigma,n)$. However,
in the case of the observable $C$ defined in Sec.~II one has to go still
farther. One has to realize that $B$ contains two steps, the selection of the
clusters in which the values of the variables are to be changed and the actual
change of the values. This is described by the decomposition
\be
B(n,\sigma;\sigma')=\sum_{\{z\}} F(n,z) G(n,z,\sigma,\sigma')
\label{3e20}
\ee
where {\{z\}} denotes the partitions of the clusters in two sets, the set in
which the spins are flipped and the one in which they are kept.

In this way one arrives at observables with the general dependences
$f(n,z,\sigma)$ and a more general joint distribution $\mu_C(n,z,\sigma)=
\mu(\sigma)A(\sigma;n)F(n,z)$. The transition probability then becomes
\be
W_C(n,z,\sigma,n',z',\sigma')=G(n,z,\sigma,\sigma')A(\sigma';n')F(n',z') \quad
{}.
\label{3e21}
\ee
The conditions providing detailed balance for $W(\sigma;\sigma')$ and
$\mu(\sigma)$ now only lead to the stationarity relation
\be
\sum_{\{n\},\{z\},\{\sigma\}}\mu_C(n,z,\sigma)W_C(n,z,\sigma,n',z',\sigma')=
\mu_C(n',z',\sigma') \quad ,
\label{3e22}
\ee
however, in general not to detailed balance in the generalized scheme.

Instead of (\ref{4e54}) and (\ref{4e55}) one now has to deal with
\be
\langle f_t\rangle=\sum_{\{n\},\{z\},\{\sigma\}}\mu_C(n,z,\sigma)f(n,z,\sigma)
\label{4e3}
\ee
and
\be
\langle f_t f_{t+1}\rangle=\sum_{\{n\},\{z\},\{\sigma\},\{n'\},\{z'\},
\{\sigma'\}}
\mu_C(n,z,\sigma)f(n,z,\sigma)W_C(n,z,\sigma,n',z',\sigma')f(n',z',\sigma')
\label{3e22a}
\ee
and to discuss spectral properties on the basis of the more general scalar
product
\be
(f,g)_z= \sum_{\{n\},\{z\},\{\sigma\}}\mu_C(n,z,\sigma)
f^*(n,z,\sigma)g(n,z,\sigma) \quad .
\label{3e24}
\ee
In the special case where $f(n,z,\sigma)=f(\sigma)$ from (\ref{4e3}) and
(\ref{3e22a}) one gets back to (\ref{4e54}) and (\ref{4e55}), respectively.
If $f(n,z,\sigma)=f(n,z)$, as holds for the observable $C$, one obtains
$\langle f_t\rangle=\sum_{\{n\}\{z\}} \tilde{\mu}_C(n,z)f(n,z)$ with
$\tilde{\mu}_C(n,z)=\tilde{\mu}(n)F(n,z)$ for (\ref{4e3}), however, for
(\ref{3e22a}) in general no simplification occurs.

Thus one is confronted with the situation where there is no detailed balance
but only stationarity and where the transition operator is not self-adjoint.
The projections in the spectral representation then are described by a
biorthogonal system and need not to be self-adjoint. Then the
weights of the eigenvalues must no longer be positive. This explains the
occurrence of negative contributions for the observable $C$ which have been
identified in the simulations and determined with some precision for SC in
$d=2$.

In the case of complex eigenvalues one has to note that they come in conjugate
complex pairs because the transition matrix is real. For such a pair, being
associated to a pair of complex conjugate projection matrices, the contribution
to (\ref{4e11}) is of the form $c\lambda^t+c^{*}\lambda^{*t}$. Parametrizing by
$\lambda= e^{-1/\tau}e^{i\phi}$ and $c=|c|e^{i\psi}$, the time dependence
\be
e^{-t/\tau}\mbox{cos}(\phi t+\psi)
\label{4e12}
\ee
is obtained. The oscillating behaviors which in the case of the observable $C$
for SC in $d=2$ are seen on smaller lattices are consistent with the form which
one gets if $\phi$ and $\psi$ are not too far from $\pi$. Such types of
behaviors are generally noticed for SC and LC, though with superimposed
decaying
modes and with larger errors in other cases.

\vspace{1.0cm}

{\Large \bf Acknowledgements}

This work has been supported in part by the Deutsche Forschungsgemeinschaft
through grant Ke 250/7-1.
The computations have been done on the CONVEX C230 of Marburg University.

\newpage
\vspace{2.0cm}

\renewcommand{\baselinestretch}{1.6}

\newpage

\renewcommand{\baselinestretch}{2.0}
\small\normalsize

{\Large \bf Figure Captions}

\begin{tabular}{rl}
FIG. 1. & $\langle C\rangle_{LC}$ and the difference
          $\langle C\rangle_{SC}- \langle C\rangle_{LC}$ denoted by DC\\
        & in $d=2$, 3, and 4 dimensions as functions of $L$.\\
FIG. 2. & $\tau$, $\tau_{int}^E$, $\tau_{int}^C$, $\tau_{int}^{\chi}$, and
          $\tau_{int}^M$
          for SC in $d=2$ as functions of $L$.\\
FIG. 3. & $\tau_{int}^E$, $\tau_{int}^C$, $\tau_{int}^{\chi}$, and
          $\tau_{int}^M$
          for LC in $d=2$ as functions of $L$.\\
FIG. 4. & $\tau$, $\tau_2$, and $\tau_{int}^E$, $\tau_{int}^C$,
          $\tau_{int}^{\chi}$, $\tau_{int}^M$
          for WO in $d=2$ as functions of $L$.\\
FIG. 5. & $c^E$, $c^{\chi}$, and $c^M$
          for SW in $d=2$ as functions of $L$.\\
FIG. 6. & $c^E$, $c^{\chi}$, and $c^E_2$, $c^{\chi}_2$
          for LC in $d=4$ as functions of $L$.\\
FIG. 7. & $c^E$, $c^{\chi}$, and $c^E_2$, $c^{\chi}_2$
          for WO in $d=3$ as functions of $L$.\\
\end{tabular}

\vspace{3.0cm}

\renewcommand{\baselinestretch}{1.3}
\small\normalsize
\begin{center}

TABLE 1

Fit to $kL^z$ for $\langle C\rangle$.

\begin{tabular}{|c|c|c|c|c|}\hline
     & \multicolumn{2}{c|}{LC}  & \multicolumn{2}{c|}{WO} \\ \cline{2-5}
$d$ &      $z$     &   $k$      & $z$     & $k$ \\ \hline
   2 &  -0.124(1) &  1.004(2)   &  -0.248(2) &  1.08(2)   \\
   3 &  -0.50(2)  &  1.05(3)    &  -1.005(5) &  1.39(2)   \\
   4 &  -0.90(2)  &  1.08(3)    &  -1.80(4)  &  1.65(12)  \\ \hline
\end{tabular}

\newpage

TABLE 2

Fit to $kL^z$ for SC and SW.

\begin{tabular}{|c|l|c|c|c|c|}\hline
   & & \multicolumn{2}{c|}{SC}  & \multicolumn{2}{c|}{SW} \\ \cline{3-6}
$d$ &                     &   $z$     &   $k$      & $z$     & $k$ \\ \hline
 4 & $\tau$               &  0.87(2)  &  0.42(2)   &  0.82(3)  &  0.86(4)  \\
   & $\tau_{int}^E$       &  0.82(2)  &  0.46(2)   &  0.75(3)  &  0.93(4)  \\
   & $\tau_{int}^{\chi}$  &  0.88(1)  &  0.38(1)   &  0.80(2)  &  0.87(3)  \\
   & $\tau_{int}^M$       &  0.88(2)  &  0.34(2)   &  0.77(2)  &  0.85(2)  \\
   & $c^E$                & -0.03(3)  &  1.04(5)   & -0.06(2)  &  1.05(2)  \\
   & $c^{\chi}$           &  0.08(4)  &  0.76(6)   & -0.03(2)  &  1.01(3)  \\
   & $c^M$                &  0.08(6)  &  0.65(8)   & -0.04(3)  &  0.94(6)  \\
\hline
 3 & $\tau$               &  0.71(2)  &  0.44(2)   &  0.61(2)  &  1.04(6)  \\
   & $\tau_{int}^E$       &  0.68(2)  &  0.47(2)   &  0.58(2)  &  1.09(4)  \\
   & $\tau_{int}^{\chi}$  &  0.67(2)  &  0.43(2)   &  0.57(2)  &  1.10(4)  \\
   & $\tau_{int}^M$       &  0.67(2)  &  0.40(2)   &  0.55(2)  &  1.07(4)  \\
   & $c^E$                & -0.01(1)  &  1.00(2)   & -0.02(1)  &  1.01(2)  \\
   & $c^{\chi}$           & -0.00(2)  &  0.86(4)   & -0.05(1)  &  1.04(2)  \\
   & $c^M$                &  0.01(2)  &  0.75(4)   & -0.07(2)  &  1.00(4)  \\
\hline
 2 & $\tau$               &  $\cap$   &  $\cap$    &  0.34(2)  &  1.27(3)  \\
   & $\tau_{int}^E$       &  $\cap$   &  $\cap$    &  0.33(3)  &  1.31(5)  \\
   & exp$(\tau_{int}^{\chi})$ & 0.47(2) & 1.13(4)  &  0.84(2)  &  2.21(11) \\
   & exp$(\tau_{int}^M)$  &  0.42(1)  &  1.17(2)   &  0.75(2)  &  2.44(9)  \\
   & $c^E$                & -0.00(1)  &  0.98(1)   & -0.01(1)  &  1.00(2)  \\
   & $c^{\chi}$           & -0.05(3)  &  0.94(9)   &  $\cap$   &  $\cap$   \\
   & $c^M$                & -0.07(2)  &  0.89(5)   &  $\cap$   &  $\cap$   \\
   & $c^C$                & -0.23(2)  &  2.22(8)   &           &     \\ \hline
\end{tabular}

\newpage

TABLE 3

Fit to $kL^z$ for LC and WO.

\begin{tabular}{|c|l|c|c|c|c|}\hline
   & & \multicolumn{2}{c|}{LC}  & \multicolumn{2}{c|}{WO} \\ \cline{3-6}
$d$ &                     &   $z$     &   $k$      & $z$     & $k$ \\ \hline
 4 & $\tau$               &  0.96(12) &  0.53(12)  &  2.08(9)  &  0.30(4)  \\
   & $\tau_2$             &  0.8(2)   &  0.4(2)    &  2.07(18) &  0.13(4)  \\
   & $\tau_{int}^E$       &  1.08(2)  &  0.36(2)   &  1.95(3)  &  0.31(2)  \\
   & $\tau_{int}^{\chi}$  &  0.97(2)  &  0.34(2)   &  1.80(3)  &  0.32(2)  \\
   & $\tau_{int}^M$       &  0.97(3)  &  0.31(2)   &  1.79(3)  &  0.34(2)  \\
   & $\tau_{int}^C$       &  0.82(1)  &  0.56(2)   &  0.6(2)   &  0.6(2)   \\
   & $c^E+c^E_2$          &  0.03(2)  &  0.93(3)   &  0.00(1)  &  1.00(1)  \\
   & $c^{\chi}+c^{\chi}_2$&  0.06(8)  &  0.74(12)  & -0.07(15) &  1.1(2)   \\
\hline
 3 & $\tau$               &  0.83(7)  &  0.44(6)   &  1.35(3)  &  0.42(2)  \\
   & $\tau^E_2$           &  0.8(3)   &  0.23(9)   &  1.17(9)  &  0.25(5)  \\
   & $\tau_{int}^E$       &  0.87(1)  &  0.38(1)   &  1.36(1)  &  0.36(1)  \\
   & $\tau_{int}^{\chi}$  &  0.75(1)  &  0.39(1)   &  1.22(2)  &  0.39(1)  \\
   & $\tau_{int}^M$       &  0.76(2)  &  0.36(1)   &  1.21(2)  &  0.41(2)  \\
   & $\tau_{int}^C$       &  0.66(2)  &  0.60(2)   &  0.36(3)  &  0.75(5)  \\
   & $c^E+c^E_2$          &  0.01(1)  &  0.96(2)   &  0.00(1)  &  1.00(1)  \\
   & $c^{\chi}+c^{\chi}_2$&  0.2(2)   &  0.5(2)    & -0.10(4)  &  1.17(9)  \\
   & $c^E$                &  0.13(5)  &  0.63(8)   &  0.11(6)  &  0.61(8)  \\
\hline
 2 & $\tau$               &  0.46(2)  &  0.50(3)   &  $\cap$   &  $\cap$   \\
   & $\tau_{int}^C$       &  $\cap$   &  $\cap$    &  0.09(1)  &  0.89(2)  \\
   & $c^E+c^E_2$          &           &            &  0.00(1)  &  1.00(1)  \\
   & $c^{\chi}+c^{\chi}_2$&           &            & -0.03(3)  &  0.98(2)  \\
   & $c^E$                &  0.06(3)  &  0.73(7)   &  0.05(4)  &  0.67(8)  \\
   & $c^{\chi}$           & -0.03(3)  &  0.65(8)   & -0.01(9)  &  0.54(18) \\
   & $c^M$                & -0.04(4)  &  0.62(8)   & -0.09(9)  &  0.7(2)   \\
   & $c^C$                & -0.11(3)  &  1.19(12)  & -0.8(2)   &  2.1(5)   \\
   & $c^E_2$              &           &            & -0.15(10) &  0.35(11) \\
   & $c^{\chi}_2$         &           &            & -0.07(7)  &  0.46(12) \\
\hline
\end{tabular}

\end{center}

\end{document}